\documentclass[aoas,MSNbibl,nameyear,seceqn,rotating,dvips]{arximspdf}
\usepackage{dcolumn}
\usepackage{graphicx}

\doi{10.1214/12-AOAS599} 
\volume{7}
\issue{1}
\pubyear{2013}
\firstpage{543}
\lastpage{569}

\makeatletter

\newcommand{\bott}{\,\bot\,}

\newcolumntype{k}[1]{D{,}{}{#1}}
\newcolumntype{d}[1]{D{.}{.}{#1}}

\newproclaim{remark}{Remark}

\newcommand{\ubeta}{\bolds{\beta}}
\newcommand{\utheta}{\bolds{\theta}}
\newcommand{\ulambda}{\bolds{\lambda}}
\newcommand{\unu}{\bolds{\nu}}
\newcommand{\upsi}{\bolds{\psi}}
\newcommand{\upi}{\bolds{\pi}}
\newcommand{\uE}{\mathbf{E}}
\newcommand{\uf}{\mathbf{f}}
\newcommand{\uG}{\mathbf{G}}
\newcommand{\uS}{\mathbf{S}}
\newcommand{\uV}{\mathbf{V}}
\newcommand{\uW}{\mathbf{W}}

\let\sv@tabnotetext\tabnotetext
\let\sv@tabnotemark@fmt\tabnotemark@fmt
\long\def\legend#1{{\let\tabnote@indent\leavevmode\sv@tabnotetext[]{}{#1}}}

\makeatother

\begin{document}
\begin{frontmatter}

\title{Bayesian semiparametric analysis for two-phase studies of
gene-environment
interaction\thanksref{T0}}
\runtitle{Two-phase studies of gene-environment interaction}

\thankstext{T0}{Genotyping and data collection were
supported by R01 CA81488 and N01 CN43302.}

\begin{aug}
\author[A]{\fnms{Jaeil}~\snm{Ahn}\corref{}\thanksref{t1}\ead[label=e1]{jaeil@umich.edu}},
\author[B]{\fnms{Bhramar}~\snm{Mukherjee}\thanksref{t1,t2,t3}\ead[label=e2]{bhramar@umich.edu}},
\author[C]{\fnms{Stephen~B.}~\snm{Gruber}\thanksref{t3}\ead[label=e3]{sgruber@usc.edu}}
\and
\author[D]{\fnms{Malay}~\snm{Ghosh}\thanksref{t1}\ead[label=e4]{ghoshm@stat.ufl.edu}}
\runauthor{Ahn, Mukherjee, Gruber and Ghosh}
\affiliation{University of Michigan, University of Michigan,\break
University of Southern California and University of Florida}
\address[A]{J. Ahn\\
Department of Biostatistics\\
University of Michigan\\
M4153 SPH II \\
1415 Washington Heights \\
Ann Arbor, Michigan 48109-2029\\
USA \\
\printead{e1}}
\address[B]{B. Mukherjee\\
Department of Biostatistics\\
University of Michigan\\
M4166 SPH II \\
1415 Washington Heights \\
Ann Arbor, Michigan 48109-2029\\
USA \\
\printead{e2}}
\address[C]{S. B. Gruber\\
USC Norris Comprehensive Cancer Center\\
University of Southern California\\
1441 Eastlake Ave, NOR 8302L\\
Los Angeles, California 90089-9181 \\
USA \\
\printead{e3}}
\address[D]{M. Ghosh\\
Department of Statistics\\
University of Florida\\
P.O. Box 118545 \\
Gainesville, Florida 32611-8545\hspace*{2.1pt} \\
USA \\
\printead{e4}} 
\end{aug}

\thankstext{t1}{Supported in part by NSF Grant DMS-10-07494.}

\thankstext{t2}{Supported by R03 CA156608 and NIH/NIEHS Grant
ES020811.}

\thankstext{t3}{Supported by NIH Grant U19 NCI-895700.}

\received{\smonth{8} \syear{2011}}
\revised{\smonth{6} \syear{2012}}

%
\begin{abstract}
The two-phase sampling design is a cost-efficient way of collecting
expensive covariate information on a judiciously selected subsample. It
is natural to apply such a strategy for collecting genetic data in a
subsample enriched for exposure to environmental factors for
gene-environment interaction \mbox{($G$ x $E$)} analysis. In this paper, we
consider two-phase studies of $G$ x $E$ interaction where phase I data
are available on exposure, covariates and disease status. Stratified
sampling is done to prioritize individuals for genotyping at phase II
conditional on disease and exposure. We consider a Bayesian analysis
based on the joint retrospective likelihood of phases I and II
data. We address several important statistical issues: (i) we consider
a model with multiple genes, environmental factors and their pairwise
interactions. We employ a Bayesian variable selection algorithm to
reduce the dimensionality of this potentially high-dimensional model;
(ii) we use the assumption of gene--gene and gene-environment
independence to trade off between bias and efficiency for estimating
the interaction parameters through use of hierarchical priors
reflecting this assumption; (iii) we posit a flexible model for the
joint distribution of the phase I categorical variables using the
nonparametric Bayes construction of Dunson and Xing
[\textit{J. Amer. Statist. Assoc.} \textbf{104} (2009)
1042--1051]. We carry
out a small-scale simulation study to compare the proposed Bayesian
method with weighted likelihood and pseudo-likelihood methods that are
standard choices for analyzing two-phase data. The motivating example
originates from an ongoing case-control study of colorectal cancer,
where the goal is to explore the interaction between the use of statins
(a~drug used for lowering lipid levels) and 294 genetic markers in the
lipid metabolism/cholesterol synthesis pathway. The subsample of cases
and controls on which these genetic markers were measured is enriched
in terms of statin users. The example and simulation results illustrate
that the proposed Bayesian approach has a number of advantages for
characterizing joint effects of genotype and exposure over existing
alternatives and makes efficient use of all available data in both
phases.
\end{abstract}

%

\end{frontmatter}

\section{Introduction}
Case-control studies are popular analytical tools, particularly in cancer
epidemiology, for assessing gene-disease association where the
allele/genotype frequencies at a bi-allelic single nucleotide
polymorphism (SNP) locus are compared between cases and controls.
Recent genomewide case-control association studies (GWAS) have been
remarkably successful in identifying susceptibility loci for many
cancers [\citet{Yeaetal07}, \citet{Hunetal07}, \citet{Amuetal09}].
A large fraction of variability in the different cancer traits still
remain unexplained, with the identified SNPs contributing modestly to
prediction of disease risk [\citet{Wacetal10}, \citet{Paretal10}].
In search of the missing heritability, it is thus natural to study the genetic
architecture of a cancer phenotype in conjunction with the known
environmental risk factors (environmental toxins, dietary exposures,
physical activity levels, medication use and other behavioral risk
factors). In the post-GWAS era, more efficient statistical
approaches to characterize such complex gene-environment ($G$ x $E$)
interactions, in terms of both design and analytic
tools, have become a pressing need in cancer epidemiology research.

Variants of the case-control sampling design have been often
employed in epidemiologic studies. Two-phase stratified sampling
[\citet{Ney38}] is an efficient alternative to the traditional cohort
and case-control designs [\citet{Coc63}] from cost and
resource-saving perspectives. A typical application of two-phase
sampling is for collecting expensive covariate information, for
example, novel biomarkers or genotype data on a prioritized
subsample of the initial study base. In particular, we
will consider the following setup: the binary disease outcome or case-control
status $D$,
some relatively inexpensive covariates ($\uS$) and environmental
data ($E$) are collected at phase I ($P_1$). At phase II ($P_2$),
genotype data ($G$) is
collected on a subset selected from the phase $I$ sample. To select this
phase~II subsample, stratified sampling with strata defined by
phase I data ($D$, $E$ and possibly~$\uS$) is implemented.

There is a large amount of literature on two-phase designs, using
different likelihood based approaches [\citet{HorTho52},
\citet{FlaGre91}, \citet{BreCai88}] or estimating score
approaches [\citet{ReiPep95}, \citet{ChaCheBre03},
\citet{RobRotZha94}]. Maximum likelihood inference for such
problems was considered in the pioneering work of \citet{ScoWil97}
and Breslow and Holubkov (\citeyear{BreHol97N1,BreHol97N2}).
\citet{LawKalWil99} and \citet{BreCha99} compare and contrast
several approaches for analyzing two-phase data. It has been noted that
adding more phases can lead to further efficiency gains, consequently,
the two-phase design has been generalized to multi-phase designs
[\citet{WhiHal}, \citet{LeeScoWil10}].
\citet{HanChe11} propose an intermediate phase between phases I and
II to reduce participation bias caused by differential
participation.

The potential for such sampling designs for $G$ x $E$ studies has been
indicated in \citet{Dur10}. Many GWAS adopt this sampling at the
design phase, but little attention is paid at the analysis
stage to address the sampling design, thus potentially leading to biased
estimates. To the best of our knowledge, literature on two-phase
studies of $G$ x $E$ interaction is very limited. \citet{ChaChe07} proposed maximum likelihood inference using a novel
regression model for \mbox{$G$ x $E$} interaction studies where
second stage sampling was carried out based on disease outcome and
family history. Asymptotic theories were established under the
assumption of independence of the genetic and environmental
factors in the population.%


Multiple papers [\citet{PieWeiTay94}, \citet{UmbWei97},
\citet{ChaCar05}] attest the phenomenon of gaining efficiency in
studies of $G$ x $E$ by exploiting independence between the genetic and
environmental factors under case-control sampling. Under such
constraints, it is beneficial to use the retrospective likelihood for
estimating interaction parameters instead of standard prospective
logistic regression. However, with departures from these constraints,
biases in estimating the interaction parameter can occur under
retrospective methods. Several researchers have addressed this issue
and proposed more robust strategies for testing $G$ x $E$ interaction
[Mukherjee et al. (\citeyear{Muketal08,Muketal10}),
\citet{MukCha08}, \citet{Vanetal08}, \citet{LiCon09},
\citet{MurLewGau09}]. There is no standard multivariate tool for
handling multiple genetic markers simultaneously for $G$ x $G$ and $G$
x $E$ studies that data-adaptively exploits gene--gene and
gene-environment independence for gaining efficiency in estimating
\textit{multiple} SNP x $E$ interaction parameters in a potentially
high-dimensional model.

Bayesian literature on two-phase studies, even beyond the context of
$G$ x $E$ studies, is also very limited. \citet{HanWak07}
presented the first hierarchical Bayesian work that closely relates to
such data structure. The Bayesian framework presented in this paper
appears to be a natural route to explore for multiple reasons. First,
Bayesian estimation can lead to efficient computational algorithms, as
the two-phase likelihood is naturally a missing data likelihood.
Second, for $G$ x $E$ studies, Bayesian methods provide data-adaptive
shrinkage to leverage the constraints of gene-environment independence
by imposing informative priors around this assumption. Third, we
incorporate Bayesian variable selection features which help us to
handle a potentially high-dimensional disease risk model with main
effects and interactions of multiple genes and environmental factors
simultaneously. Fourth, we use the clever nonparametric Bayesian
construction of \citet{DunXin09} as a substitute for profile
likelihood in the frequentist setting to construct the retrospective
likelihood under two-phase sampling. The current paper thus contributes
to analysis of $G$ x $E$ studies with multiple markers/environmental
exposures under an outcome-exposure stratified two-phase sampling
design by offering a new Bayesian treatment of the problem. Our data
analysis and simulation studies illustrate that for characterizing
subgroup effects of the environmental exposure across genotype
categories, our method provides gain in efficiency
compared to other alternatives. Moreover, there are no comparable
alternatives that can offer the flexibility of our method in terms of
multi-marker models and efficient \mbox{$G$ x $E$} analysis under the
two-phase design.

The paper is largely motivated by an example that originates from a
population based case-control study of colorectal cancer (CRC)
in Israel, namely, the Molecular Epidemiology of Colorectal Cancer
(MECC) study. Statins (our environmental factor $E$) are a class of
lipid-lowering drugs used by more than 25 million individuals
worldwide for reducing cardiovascular disease risk. The MECC study
was the first to establish a chemoprotective association of
statins with risk of CRC [\citet{Poyetal05}]. Follow-up
individual studies and a meta analysis of 18 studies have confirmed
this association [\citet{Hacetal09}].
The benefit of statins for reducing CRC risk has been shown
to vary with genetic variations in the HMGCR (3-Hydroxy-3-methylglutaryl
coenzyme A reductase) gene, a gene involved in cholesterol
synthesis [\citet{Lipetal10}]. To understand the mechanism of effect
modification further, investigators measured 294 SNPs in 40
genes, including HMGCR (our set of genetic factors $\uG$),
selected in the cholesterol synthesis/lipid metabolism pathway. The
subsample selected for genotyping from the study population of all
cases and controls was chosen by stratified sampling conditional on
statin use ($E$) and case-control status ($D$) where statin users
were purposefully oversampled. This sampling strategy was adopted
due to limited budgetary resources and DNA samples. Complete statin
use ($E$) data and other basic demographic covariates ($\uS$) were
available on the entire study base (phase I or $P_1$), and genetic
data on these 294 SNPs were only available for the phase II
subsample~($P_2$).

\begin{figure}

\includegraphics{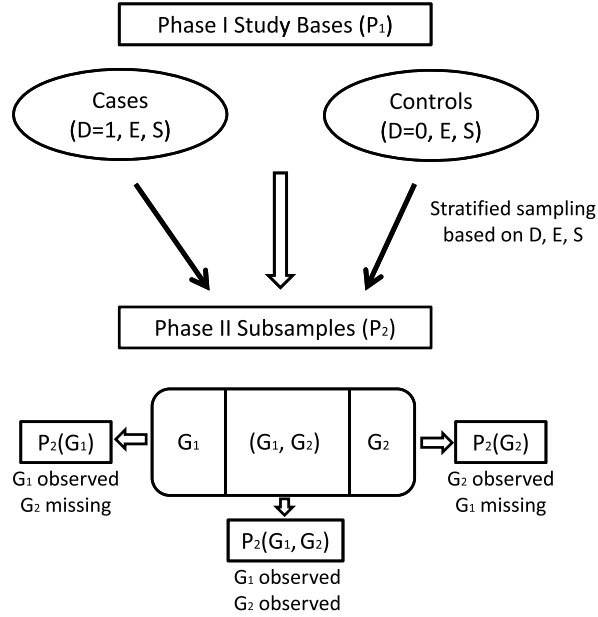}

\caption{Data structure under two-phase sampling
with partial missingness in phase II genetic covariates from the
Molecular Epidemiology of Colorectal Cancer study.}\label{tpfigure1}
\end{figure}

In addition, in the MECC study, due to experimental and laboratory logistics,
genotype data were missing on a subset of individuals selected in
$P_2$ on a group of genes ($\uG_1$, say) and on a different subset
of individuals on another group of genes ($\uG_2$, say). This led to
a nonmonotone missing data structure with some individuals in $P_2$
having observations on both $(\uG_1,\uG_2)$ [subset denoted by
$P_2(\uG_1,\uG_2)$] and some only on $\uG_1$ [subset denoted by
$P_2(\uG_1)$] and some only on $\uG_2$ [subset denoted by
$P_2(\uG_2)$]. Figure~\ref{tpfigure1} is a flow diagram of
the sampling scheme and missingness pattern in the data.

The rest of the paper is organized as follows. In Section \ref
{tpprop} we
present the model ingredients: the likelihood, priors and
posteriors. In Section~\ref{tpexam} we discuss the analysis of statin x
gene interaction in the MECC study. In Section~\ref{tpsim} we conduct a
simulation study to compare the various maximum likelihood and score
based approaches with the Bayesian approach. Section~\ref{tpdiscuss} concludes
with a discussion.

\section{Proposed methods}\label{tpprop}
\subsection{The likelihood}
We refer to Figure~\ref{tpfigure1} for understanding the data structure
and construction of our likelihood. Let $u$ and $D$
denote the subject indicator and disease status, respectively. Here,
$E$ is environmental exposure and $\uS$ are basic demographic
covariates as described before. Let $W=(E,\uS)$. There are $N$
individuals in phase I and $M$ individuals in phase II. To simplify
notation, we write the retrospective likelihood corresponding to a
two-gene model ($G_1$, $G_2$), with the understanding that the
methods/notation can be directly extended to gene-sets ($\uG_1, \uG_2$)
where each contain multiple SNPs. The two-phase likelihood has the
following form to capture the sampling phases and the missingness
patterns in $\uG$ (Figure~\ref{tpfigure1}):
\begin{eqnarray*}
{\mathrm L}^{\mathrm{TP}} &=& \prod_{u \in P_1\setminus P_2}
\mathrm{P}(W_u|D_u) \times \prod
_{u \in P_2(G_1)} \mathrm{P}(G_{1u}, W_u|D_u)
\\
&&{}\times \prod_{u \in P_2(G_2)} \mathrm{P}(G_{2u},
W_u|D_u) \times \prod_{u \in P_2(G_1,G_2)}
\mathrm{P}(G_{1u},G_{2u},W_u|D_u).
\end{eqnarray*}

Each term in ${\mathrm L}^{\mathrm{TP}}$ can be factorized by using
$\mathrm{P}(G_1,G_2,W|D)=\{\mathrm{P}(D|G_1,\break G_2,W)$
$\mathrm{P}(G_1, G_2|W) \mathrm{P}(W)\}/\mathrm{P}(D)$. This
retrospective likelihood is then mar\-ginalized over the missing data in
each term. We assume missing completely at random
[\citet{LitRub02}] for the genotype data
collected at phase~II. The likelihood is then expressed as
%
\begin{eqnarray}
\label{tplikeli}\qquad {\mathrm L}^{\mathrm{TP}} &=& \prod
_{u \in P_1\setminus P_2} \sum_{g_1,g_2}
\mathrm{P}(D_u|g_1,g_2,W_u){
\mathrm{P}(g_1,g_2|W_u)}
\mathrm{P}(W_u) /\mathrm{P}(D_u)
\nonumber
\\
&&{}\times \prod_{u \in P_2(G_1)}\sum
_{g_2} \mathrm{P}(D_u|G_{1u},g_2,W_u){
\mathrm{P}(G_{1u},g_2|W_u)}
\mathrm{P}(W_u)/\mathrm{P}(D_u)
\\
&&{}\times \prod_{u \in P_2(G_2)}\sum
_{g_1} \mathrm{P}(D_u|g_1,G_{2u},W_u){
\mathrm{P}(g_1,G_{2u}|W_u)}
\mathrm{P}(W_u)/ \mathrm{P}(D_u)
\nonumber
\\
&&{}\times \prod_{u \in P_2(G_{1},G_{2})} \mathrm{P}(D_u|G_{1u},G_{2u},W_u){
\mathrm{P}(G_{1u},G_{2u}|W_u)}
\mathrm{P}(W_u)/\mathrm{P}(D_u),
\nonumber
\end{eqnarray}
where $\mathrm{P}(D_u)=\sum_{g_1,g_2} \int_{w}
\mathrm{P}(D_u|g_1,g_2,w)\mathrm{P}(g_1,g_2|w)\mathrm{P}(dw)$ with the integral
replaced by the sum when components of $W$ are discrete. Corresponding
to this
likelihood, there are three model ingredients:\vspace*{6pt}

1. \textsc{A disease risk model.} We assume
$\mathrm{P}(D=1|G_1=g_1,G_2=g_2,W=w;\ubeta)
=H[\{\beta_0+m(g_1,g_2,w;\ubeta)\}]$, where $H$ is the logistic
function $H(u)=\{1+\exp(-u)\}^{-1}$. Typical choice of $m$ involves,
say, for two genes $G_1$ and $G_2$,
$m(g_1,g_2,w;\ubeta)=\beta_{G_1}g_1+\beta_{G_2}g_2+\beta_{E}e+
{\ubeta^{\top}_S} \mathbf{s}+\beta_{G_1G_2} g_1g_2+ \beta_{G_1E}g_1e+\beta_{G_2E}g_2e$,
noting that $w=(e,\mathbf{s})$.\vspace*{6pt}

2. \textsc{A model for $(G_1,G_2|W=(E,\uS))$.} For
genotype data at a bi-allelic locus, $G_j$ can take three possible
values (``$g_0=\mbox{aa}$,'' ``$g_1=\mbox{Aa}$'' and ``$g_2=\mbox{AA}$''). We assume,
$\mathrm{P}(G_1=g_j,G_2=g_j'|W=w;\ulambda)=q_{jj'}(w;\ulambda), j,j'=0,1,2$.
This specification will require a joint model for multivariate
categorical data (trinary for SNP data at a bi-allelic locus). Under
gene--gene and gene-environment independence, the model
can in general be factorized conditional on covariates $\uS$, for
$j,j'=0,1,2$,
{\fontsize{10.65pt}{11pt}\selectfont{
\[
\underbrace{\mathrm{P}\bigl(G_1=g_j,G_2=g_j'|E=e,
\uS=\mathbf{s};\ulambda\bigr)=\mathrm{P}(G_1=g_j|\uS=
\mathbf{s}, \ulambda_1)\mathrm{P}\bigl(G_2=g_j'|
\uS=\mathbf{s}, \ulambda_2\bigr)}_{\mathrm{under}\ \mathrm{G}\mbox{-}\mathrm{G}\ \mathrm{and}\ \mathrm{G}\mbox{-}\mathrm{E}
\ \mathrm{independence}}.
\]}}
\noindent Instead of the above fully nonparametric model, we explore a parametric
model for
the joint distribution $\mathrm{P}(\uG_1,\uG_2|W)$. We consider a class of
log-linear models with linear by linear structure [\citet{Agr02}] for parsimonious modeling of the $(G_1,G_2|W)$ associations,
%
\begin{eqnarray}
\label{tploglinear} && \log\bigl\{\mu\bigl(G_1=g_j,
G_2=g_j'|E=e,\uS=\mathbf{s}; \ulambda\bigr)
\bigr\}
\nonumber
\\
& &\qquad =\lambda_0+ \lambda_{G_1} g_j +
\lambda_{G_2} g_j'+ \lambda_E e +
{\ulambda^{\top}_S} \mathbf{s}
\\
&&\qquad\quad{} +\lambda_{G_1G_2} g_{j} g_{j'} +
\lambda_{G_1E} g_{j} e + \lambda_{
G_2E}g_{j'}
e+ {\ulambda^{\top}_{G_1S}} g_{j} \mathbf{s}+
\ulambda^{\top}_{G_2S} g_{j'} \mathbf{s},
\nonumber
\end{eqnarray}
where $g_{j}$ are chosen ordinal scores, typically 0, 1, 2
[\citet{Agr02}]. This is the common allelic dosage coding under a
log-additive genetic susceptibility model. Our method could easily be
extended to a co-dominant coding of the genetic factor using two dummy
variables. Since log-additivity is often assumed for screening
interactions, and for simplicity of presentation in terms of one
parameter estimate as opposed to two, we proceed with
this additive coding. Additionally, even if the
true genetic susceptibility model is co-dominant with the
disease-causing allele, for a tagging marker which is correlated to
this causal allele, one would not a'priori know the direction of
association of the marker allele and causal allele.
\citet{PfeGai03} show that the additive scores are more robust to
choice of marker allele and varying correlation scenarios. In case of
high-dimensional~$\uG$, we can further reduce the dimensionality of the
problem by assuming common association parameters
$\lambda_{\mathit{GE}}$ and $\lambda_{\mathit{GS}}$ between similar
functional groups of SNPs. As discussed in \citet{Agr02}, this
Poisson log-linear model has a corresponding multinomial
representation. Thus, the probability of $P_{G_1,G_2}({g_j,
g_j'|\ulambda})=P(G_1=g_j, G_2=g_j'|E=e,\uS= \mathbf{s})$ can be
written in terms of the multinomial probabilities,
\begin{eqnarray*}
&& P_{G_1,G_2}\bigl({g_j, g_j'|
\ulambda}\bigr)
\\
&&\qquad= \exp\bigl(\lambda_{ G_1} g_j +
\lambda_{ G_2} g_j' +\lambda_{ G_1G_2}g_j
g_{j'} \\
&&\qquad\quad\hspace*{17.5pt}{} + \lambda_{ G_1E} g_j e + \lambda_{
G_2E} g_j' e + {\ulambda^{\top}_{ G_1S}} g_j \mathbf{s}+
{\ulambda^{\top}_{ G_2S}} g_j' \mathbf{s}\bigr)\\
&&\qquad\quad{}\times \Biggl(\sum_{l=0}^{2} \sum_{l'=0}^2
\exp\bigl(\lambda_{ G_1} g_l + \lambda_{ G_2}
g_l'+\lambda_{ G_1G_2}g_l g'_{l'} \\
&&\qquad\quad\hspace*{71pt}{} +
\lambda_{ G_1E} g_l e + \lambda_{ G_2E}
g'_{l'} e + {\ulambda^{\top}_{ G_1S}} g_l \mathbf{s}+
{\ulambda^{\top}_{
G_2S}} g'_{l'} \mathbf{s}\bigr)\Biggr)^{-1}.
\end{eqnarray*}

Note that gene--gene and gene-environment independence in the above
model (\ref{tploglinear}) will imply
$\lambda_{G_1E}\equiv\lambda_{G_2E} \equiv\lambda_{G_1G_2}
\equiv0$.\vspace*{6pt}

3. \textsc{A model for $W=(E, \uS)$.} A nonparametric and flexible
model for the distribution of $W$ is desired. Recall that $W$ can be a
mixed set of quantitative and categorical variables.\vadjust{\goodbreak} For the MECC
example $W$ is a set of categorical covariates, which will be our
primary focus in this paper. The approach for modeling the joint
distribution of a set of categorical variables that we follow for $W$
can also be applied to the the joint distribution of the trinary
genotype variables $\uG_1$ and $\uG_2$ in (\ref{tploglinear}) as well.
However, reflecting prior faith on the gene--gene and gene-environment
independence assumptions through direct priors on parameters
$\lambda_{G_1E}, \lambda_{G_2E},\lambda_{G_1G_2}$ in the log-linear
model is more straightforward for a practitioner (\ref{tploglinear}).
This is the primary reason for using (\ref{tploglinear}) for the second
component $P(G_1,G_2|W=(E,\uS))$.

Let $\uW_u =(E_u, \uS_u)$ denote the $W$ data corresponding to subject
$u$, $u=1,\ldots, N$. Here $W_u$ is $p \times1 $ vector of $p$
categorical variables, that is, $W_u=(w_{u1},\ldots, w_{up})$ for a
subject $u$. Assume that the $j$th component of $W$ can have $d_j$
values $j=1, \ldots, p$. In order to parsimoniously model this $(d_1
\times d_2 \times\cdots\times d_p)$ joint distribution, DX first note
that the joint distribution of two categorical variables can always be
expressed as a finite mixture of product-multinomial distributions.
Extending this idea, DX introduce a latent class index variable $z_u
\in \{1,\ldots, k\}$, such that $w_{ur}, w_{ut}, r, t \in\{1,\ldots,
p\}, r \neq t$, are conditionally independent given
$z_u$. Then the joint distribution for $\mathbf{w}_u$ has this finite
mixture representation,
%
\begin{eqnarray}
\label{tpEbasic}
&&P_W(w_{u1}=c_1, \ldots,
w_{up}=c_p)\nonumber\\
&&\qquad= \sum_{h=1}^{k}
P(w_{u1}=c_1, \ldots, w_{up}=c_p|z_u=h)P(z_u=h)
\\
&&\qquad= \sum_{h=1}^{k} P(z_u=h)
\prod_{j=1}^p P(w_{uj}=c_j|z_u=h).
\nonumber
\end{eqnarray}
For notational convenience, we rewrite (\ref{tpEbasic}) as
%
\begin{eqnarray}
\label{Dunson}
P_W(w_{u1}=c_1, \ldots,
w_{up}=c_p) &=& \pi_{c_1\cdots c_p} = \sum
_{h=1}^{k} \nu_h \prod
_{j=1}^p \psi_{h c_j}^{(j)},\nonumber\\[-8pt]\\[-8pt]
\sum
_{c_1=1}^{d_1}\cdots\sum
_{c_p=1}^{d_p}\pi_{c_1\cdots c_p}&=&1,\nonumber
\end{eqnarray}
where $\unu=(\nu_1,\ldots, \nu_k)^{\top}$ is a probability vector
with $\nu_h =P(z_u=h)$ and $\psi_{h c_j}^{(j)}=P(w_{uj}=c_j|z_u=h)$
is a $d_j \times1$
probability vector, that is, the conditional probability of
$w_{uj}=c_j$, given that subject $u$ is in latent class $h$
for $j=1,\ldots, p$. We will
discuss the choice of $k$ through a Dirichlet process prior
structure on this latent class probability model in the next
section.
%
\begin{remark}\label{remark1}
While \citet{ChaChe07} and \citet{ChaCar05} use profile likelihood for
handling the distribution of $W$ nonparametrically, it has been a
challenging task in the Bayesian framework to posit a flexible model
for $\uW=(E,\uS)$ which could be a mixture of categorical and
continuous covariates. In this mixed case, \citet{Muletal99}
model the joint distribution of the continuous covariates through a
Dirichlet process mixture of normals. Then, conditional on the
continuous covariates, the categorical variables have a joint
multivariate probit distribution.
A recent paper by \citet{BhaDun}
extends the above DX construction for categorical data to handle
joint distribution modeling of more complex data, including
continuous and discrete data. They extend the conditional
independence idea and replace the product-multinomial structure in
(\ref{Dunson}) by a product of various kernels, such as Gaussian,
Poisson and more complex univariate or multivariate distributional
kernels.
The MECC example does not require going
beyond the original DX construction, but with continuous~$E$,
this is what we would adopt.
\end{remark}
%
\begin{remark}\label{remark2}
If the phase I sample is a cohort study, with disease endpoint~$D$,
then the corresponding likelihood is proportional to
%
\begin{eqnarray}
\label{cohorttplikeli} \mathrm{L}^{\mathrm{cohort}, \mathrm{TP}} &\propto& \prod
_{u \in P_1\setminus P_2} \sum_{g_1,g_2}
\mathrm{P}(D_u|g_1,g_2,W_u){
\mathrm{P}(g_1,g_2|W_u)}
\nonumber
\\
&&{} \times \prod_{u \in P_2(G_1)}\sum
_{g_2} \mathrm{P}(D_u|G_{1u},g_2,W_u){
\mathrm{P}(G_{1u},g_2|W_u)}
\nonumber\\[-8pt]\\[-8pt]
&&{}\times \prod_{u \in P_2(G_2)}\sum
_{g_1}\mathrm{P}(D_u|g_1,G_{2u},W_u){
\mathrm{P}(g_1,G_{2u}|W_u)}
\nonumber
\\
&&{} \times \prod_{u \in P_2(G_{1},G_{2})}\mathrm{P}(D_u|G_{1u},G_{2u},W_u){
\mathrm{P}(G_{1u},G_{2u}|W_u)}.
\nonumber
\end{eqnarray}
Similarly, if environmental data $E$ is collected in phase II as well,
the first term
representing the phase I cohort likelihood can also involve an integral
over the missing $E$ data with
respect to a probability distribution $dF(E)$, exactly as in equation
(3) of \citet{ChaChe07}.
A surrogate measure of $E$, namely, $E^{*}$, may be available in phase
I and a measurement error model
relating $E$ and $E^{*}$ can also be used to construct a joint
likelihood of phases I and II data.
\end{remark}

\subsection{Priors}
As mentioned before, for this complex retrospective likelihood
formulation, we have three sets of parameters from the above three
ingredients of the likelihood. For $\ubeta$ in the disease risk model,
we use a spike
and slab type mixture prior to handle variable
selection in a high-dimensional disease risk model with multiple
markers. For $\ulambda$ in the multivariate gene model,
the Bayesian hierarchical approach provides
a flexible way to allow for uncertainty around the assumption of
gene--gene and gene-environment independence, through prior on\vadjust{\goodbreak}
$\lambda_{G_1G_2}, \lambda_{G_1E}$ and $\lambda_{G_2E}$.
When sparsity occurs in
a certain configuration of $(\uG_1,\uG_2, \uW)$ or dimension of
$(\uG_1,\uG_2, \uW)$ grows, the frequentist profile likelihood
estimation may become unstable and the log-linear model with shared
parameters across gene-sets and the DX latent mixture construction
aid with such situations.
We follow the same
sequence as in the previous section to describe the prior structure on
the parameters.\looseness=-1

1. In the presence of multiple genes in
$\uG_1$ and $\uG_2$, the logistic disease risk model can potentially
have many pairwise and higher order interaction terms. We implement
a scalable variable selection framework via spike and slab type
priors [\citet{MitBea88}, \citet{GeoMcc93}]
on the parameters $\ubeta$ in the disease risk model
$P(D|\uG_1,\uG_2,W; \ubeta)$. We impose mixture prior distributions
on each component of $\ubeta$, say, $(\beta_0, \beta_{G_1},\break
\beta_{G_2}, \beta_{E}, \beta_S, \beta_{G_1G_2}, \beta_{G_1E}$,
$\beta_{G_2E})$ for a two-gene model. In general, we denote this
vector by $\ubeta_{n_{\beta} \times1} = \{\beta_r, r=1,\ldots,
n_{\beta}\}$. Given a latent variable $p_0$ representing the mixture
weight on the ``not informative'' regression coefficients, we describe
the hierarchical prior structure as follows:
\begin{eqnarray*}
\beta_{r}|f_r, {\tau}_r &\stackrel{\mathrm{ind}} {
\sim}& N\bigl(0, f_r \tau_r^2\bigr),\qquad r=1,
\ldots, n_{\beta},
\\
f_r|v_0, p_0 &\stackrel{\mathrm{i.i.d.}}
{\sim}& p_0 \delta_{v_0}(\cdot) + (1-p_0)
\delta_{1}(\cdot),
\\
\tau_r^{-2}|a_1, a_2 &\stackrel{
\mathrm{i.i.d.}} {\sim}& \operatorname{Gamma}(a_1, a_2),
\\
p_0 &\stackrel{\mathrm{i.i.d.}} {\sim}& \operatorname{Beta}(a, b).
\end{eqnarray*}
As discussed in \citet{IshRao03}, $v_0$ in the above
specification is assumed to be a small positive value near 0. Note that
$f_r$ can assume two values $v_0$ or 1. At each iteration of
posterior sampling, $f_r$ takes value $1$ if sampled $\beta_r$ is
significantly away
from zero, implying that the $r$th covariate is potentially
informative. Note that a key feature of this prior specification is
that the marginal prior variance of $\beta_r$ is calibrated as
$\operatorname{var}(\beta_r)= f_r \tau_r^2$ and has a bimodal distribution.
Large $\operatorname{var}(\beta_r)$ can occur when $f_r=1$ and $\tau_r^2$ is large,
inducing large values of $\beta_r$, identifying potentially
informative covariates. Small values of $\operatorname{var}(\beta_r)$ occur when
$f_r$ assumes value $v_0$, leading to values
of $\beta_r$ that are near zero, suggesting that $\beta_r$ is
potentially uninformative. The value of $p_0$
controls how likely it is for $f_r$ to be $v_0$ or $1$, thus
controlling how many $\beta_r$ are nonzero
or the complexity of the model. The Gamma parameters $(a, b)$ control
the degree of parsimony through the prior on $p_0$.
We set $(a, b)=(1,1)$, that is, a uniform prior on $p_0$, for the
analysis we present in the main text.
Note that $(a_1, a_2)$ determines the prior on $\tau_r^2$ and
thus the variance of $\beta_r$. We fix $(a_1, a_2)$ at $(5, 50)$ to allow
the possibility of large prior variances on $\beta$. The values used
for the hyperparameters in the hierarchy are exactly as recommended in
\citet{IshRao03}.

2. In the joint log-linear model (\ref{tploglinear}), we typically
assume vague normal priors with large variance on the parameters
$(\lambda_{G_1}, \lambda_{G_2}, \lambda_{G_1S}$, $\lambda_{G_2S})$. In
our data example,\vadjust{\goodbreak} we have used a $N(0,10^4)$ prior. On the other hand,
for the $G$-$E$ pairwise association parameters $(\lambda_{G_1G_2},
\lambda_{G_1E}, \lambda_{G_2E})$, we reflect a priori information on
\mbox{$G$-$G$} or $G$-$E$ independence via a normal prior centered at zero but
with two different choices for the prior variance. In the first set of
priors we reflect the belief that with 95\% probability the association
parameter lies between $\log(0.8)$ and $\log(1.2)$. This leads to an
approximate $\mbox{SD}=0.1$ under a normal distribution and, thus, we assume an
informative prior of $N(0, 10^{-2})$. In the second choice, following
the empirical Bayes estimation of \citet{MukCha08}, we compute
association parameters\vspace*{1pt} for $G_1$-$G_2$, $G_1$-$E$, and $G_2$-$E$ in
the control subjects in the data, say, $\hat{\theta }$, and use a
data-driven prior $N(0, {\hat{\theta}^{2}})$ on $\lambda_{G_1G_2}$,
$\lambda_{G_1E}$ and $\lambda_{G_2E}$.

3. The mixture representation
in (\ref{Dunson}) requires determining the number of
latent classes $k$. Following DX, instead of selecting a fixed
$k$, a Bayesian nonparametric approach is carried out through the Dirichlet
process prior specification on $\unu$:
\begin{eqnarray*}
\upi&=&\sum_{h=1}^{\infty}
\nu_h \upsi_h,\qquad \upsi_h=
\upsi_h^{(1)} \otimes \cdots \otimes\upsi_h^{(p)},\qquad
h=1,\ldots, \infty,
\\
\upsi_{h}^{(j)} &\sim& \operatorname{Dirichlet}(a_{j1},\ldots,
a_{j d_j})\qquad\mbox{independently for } j=1,\ldots, p,
\\
\nu_h &=& \sum_{h=1}^{\infty}
V_h\prod_{l < h} (1-V_l),\qquad
V_h \sim \operatorname{Beta}(1, \alpha),
\nonumber
\\
\alpha&\sim& \operatorname{Gamma}(a_\alpha, b_\alpha),
\end{eqnarray*}
where $\otimes$ is the outer product. The parameter $\alpha$ is a
hyper-parameter that controls the rate of decrease from the
stick-breaking process [\citet{Set94}]. For example, in the case of
small values
of $\alpha$, $\nu_h$ decreases toward zero quickly with increasing
$h$, thus putting most of the weight on the first few components,
leading to a
sparse representation. The hyperprior on $\alpha$ allows one to
data-adaptively determine the degree of sparseness or the number of
components needed. As discussed in \citet{DunXin09}, we set $(a_\alpha,
b_\alpha)=(1/4, 1/4)$
for a vague prior which implies the probability of independence
across components of $w$ in the product multinomial model to be $0.5$.
We set uniform priors for each category probability $\upsi$ with
$a_{j1}=\cdots=a_{j d_j}=1$, for $j= 1,\ldots, p$ and let the data
dominate over priors. To minimize large numbers of
mixture components instead of using infinite mixtures, we truncate the
maximum of the number
of mixture components $k$ at $30$ in the real data example [\citet{Ahnetal}].
We study sensitivity with respect to this truncation threshold in Table 1.

\subsection{Posterior sampling}
In the full likelihood (\ref{tplikeli}), we would like to point out
that the three components are linked with each other through the sum
over each component in the expression for $P(D)$ in the denominator. We
denote the\vadjust{\goodbreak} two-phase likelihood in (\ref{tplikeli}) by
$\mathrm{L}_{\mathrm{TP}}$ which involves the parameters $(\ubeta,
\ulambda, \upsi, \uV, \alpha)$. The full conditionals are not reducible
to a simpler closed form and are best represented by the following
proportionality relations:
\begin{eqnarray*}
\beta_r|\cdot&\propto& {\mathrm L}^{\mathrm{TP}} \times\exp\biggl(-
\frac{\beta_r^2}{2 f_r \tau_r^2}\biggr),\qquad r=1,\ldots, n_{\beta},
\nonumber
\\
\tau_r^{-2}|\cdot&\propto& \operatorname{Gamma}
\biggl(a_1+0.5, a_2+\frac{\beta_r^2}{2 f_r}\biggr),
\\
f_r|\cdot&\propto& \bigl\{I(f_r = v_0)p_0
+ I(f_r = 1) (1-p_0) \bigr\} \times \exp\biggl(-
\frac{1}{2 f_r \tau_r^2} \beta_r^2\biggr)\times f_r^{-0.5},
\\
p_0 |\cdot&\propto& \operatorname{Beta}\Biggl(a +\sum
_{r=1}^{n_\beta}I(f_r = v_0), b
+\sum_{r=1}^{n_\beta}I(f_r = 1)
\Biggr),
\\
\lambda_l|\cdot&\propto& {\mathrm L}^{\mathrm{TP}} \times\exp
\biggl(-\frac{\lambda_l^2}{2\sigma^2}\biggr),\qquad l=1,\ldots, n_{\lambda}, 
\end{eqnarray*}
where $n_{\beta}$ and $n_{\lambda}$ again represent the number of
parameters in $(\ubeta, \ulambda)$, respectively.

\textit{Posterior sampling corresponding to $P(\uW)$}:
Let us recapitulate the model structure for $\uW$ which is
essentially a Dirichlet process mixture of discrete Dirichlet
kernels. For $u=1, \ldots, N$ and $j=1, \ldots, p$,
\begin{eqnarray*}
w_{uj} & \sim& \operatorname{Multinomial} \bigl(\{1,\ldots, d_j\},
\psi^{j}_{z_u, 1},\ldots, \psi^{j}_{z_u, d_j}
\bigr),
\\
z_u &\sim& V_h\prod_{l < h}
(1-V_l)\delta_h,\qquad V_h \sim \operatorname{Beta}(1,\alpha),\qquad
\alpha\sim \operatorname{Gamma}(a_\alpha, b_\alpha).
\end{eqnarray*}
DX present an efficient data-augmented Gibbs sampling algorithm by
augmenting the likelihood with latent constructs following \citet{Wal07}. The details of the updating steps are described
in the supplemental article [\citet{Ahnetal}].

Note that while the entire likelihood in DX is constituted of $W$ data
only, in our problem, $P(\uW)$ is embedded as a component in the joint
retrospective likelihood ${\mathrm L}^{\mathrm{TP}}$ in
(\ref{tplikeli}). Thus, for updating the parameters involved in
$\mathrm{P}(\uW)$, say, $\utheta(\mbox{$=$}\{ \upsi, \uV, \alpha\})$, we use the
Metropolis Hastings algorithm. Only the terms
$\prod_{u}\mathrm{P}(\uW_u) /\mathrm{P}(D_u)$ from the full likelihood
(\ref{tplikeli}) involve $\utheta$, where
$\mathrm{P}(D_u)=\sum_{g_1,g_2}
\sum_{\mathbf{w}}\mathrm{P}(D_u|g_1,g_2, \mathbf{w})$
$\mathrm{P}(g_1,g_2|\mathbf{w})\mathrm{P}(\mathbf{w})$. We draw
$\utheta$ following the DX algorithm and for the proposal density of
$\utheta$ we consider the implied full conditional
$q(\utheta^{\mathrm{new}}|\uW)$ as determined by this algorithm. Then given
$\ulambda, \ubeta$, we repeat the following updates of $\utheta$:
\begin{itemize}
\item[$\bullet$] At iteration $l$, sample a vector $\utheta^{\mathrm{new}}$
from $q(\utheta^{\mathrm{new}}|\uW)$
as described in the \citet{DunXin09} algorithm.
\item[$\bullet$] Compute the acceptance ratio
\[
r\bigl(\utheta^{\mathrm{new}}, \utheta_l\bigr) = \operatorname{min}
\biggl[1, \frac{\prod_{u}
\mathrm{P}(D_u|\utheta_{l}, \ulambda, \ubeta) } {\prod_{u}
\mathrm{P}(D_u|\utheta^{\mathrm{new}}, \ulambda, \ubeta) } \biggr].
\]
In calculating the acceptance ratio, we note that the numerator
and denominator $\prod_{u} \{\mathrm{P}(W_u|\utheta^{\mathrm{new}})\}p(\utheta^{\mathrm{new}})$
$q(\utheta_{l}|\uW)/ \prod_{u} \{\mathrm{P}(W_u|\utheta_{l})\}
p(\utheta_{l})q(\utheta^{\mathrm{new}}|\break\uW)$
are canceled out where $p(\utheta)$ is a prior for $\utheta$.
\item[$\bullet$] If $r(\utheta^{\mathrm{new}}, \utheta_l) < U$ where $U \sim
\operatorname{unif}(0,1)$,
we set $\utheta_{l+1}=\utheta^{\mathrm{new}}$. Otherwise, the candidate vector
$\theta^{\mathrm{new}}$
is rejected and $\utheta_{l+1}=\utheta_{l}$.
\item[$\bullet$] Repeat the steps until the posterior chains converge.
\end{itemize}

Given the full conditionals, we implement the Gibbs sampler [\citet{GemGem84}] with Metropolis Hastings updates to sample from
respective full conditional distributions. For each parameter, we
iterate 50,000 times and discard the first 40,000 iterations as
``burn-in.'' We check convergence of the chains using trace plots and
the numerical diagnostic statistic ``potential scale reduction factor''
[\citet{Dav92}] using the R package CODA [\citet{Pluetal}].
Auto and cross-correlation checks are performed and a thinning of
every tenth observation is carried out. Remaining posterior samples
are used to construct estimated posterior summaries needed for
Bayesian inference.

\section{The Molecular Epidemiology of Colorectal Cancer study}\label{tpexam}
In this section we describe the motivating example from the MECC
study in detail and present analysis results. We use data on 1746
cases and 1853 controls with completely observed response to the
question whether statins were used for more than 5 years. The binary
variable ``statin use of at least 5 years'' ($E$) is the environmental
factor of interest with 91\% ``NO'' and 9\% ``YES.''

We adjust for completely observed confounders and precision variables
($\uS$): age ($S_1$), gender
($S_2$), ethnicity ($S_3$), physical activity ($S_4$), family
history of CRC ($S_5$), vegetable consumption ($S_6$), NSAID usage
within 3 year ($S_7$) and Aspirin usage within 3 year ($S_8$). Age
and ethnicity variables were dichotomized as Age $\geq$ or $<$ 50
(94\% and 6\%, resp.), and ``Ashkenazi'' and ``Non-Ashkenazi''
(68\% and 32\%, resp.). Gender $(S_3)$ was coded as~1 (50\%)
for male and 0 (50\%) for female. The remaining binary factors
$(S_4, S_5, S_6, S_7, S_8)$ are classified to 1 or ``YES'' with the
proportions of (0.36, 0.09, 0.31, 0.02, 0.20), respectively.

For genotyping at phase II, stratified-sampling based on the disease
status ($D$) and statin use ($E$) was carried out. All case-control
subjects with statin use (``YES'') were included at the phase II sample.
We have
1200 cases and 1200 controls at phase II with data available on
294 trinary SNPs $\uG=(G_1,\ldots, G_{294})$. Genotype data are not
completely observed even at phase II due to technical genotyping
failures for a limited number of SNPs. Among 2400 case-control
subjects at phase II, 56 subjects and 20 had partial genotype information
on two subsets of SNPs. We did not have a dense set of
markers typed across the genome to successfully impute these missing
genotypes, thus we consider a marginalized likelihood as in
(\ref{tplikeli}).

Among 294 SNPs, we first illustrate our methods with two SNPs on two
genes, $RS1800775$ on CETP ($G_1$) and $RS1056836$ on CYP1B1
($G_2$), where both SNPs exhibit significant interactions with
statin use in an initial single marker interaction
analysis. We compare our methods for this simple two SNP model to some of
the alternative methods that can only handle single marker interaction
analysis. The raw frequencies of the cross-classification of
case-control status ($D$), statins~($E$), genotypes $G_1$ and $G_2$
are shown in online supplementary Table 1 [\citet{Ahnetal}]. Simple logistic
regression analysis was carried out to examine $G_1$-$E$ and
$G_2$-$E$ association among control subjects and yielded odds ratios
of 1.11 and 1.01 and corresponding p-values of 0.30 and 0.91,
respectively. Based on a chi-squared test for independence,
$G_1$-$G_2$ reveals no association (p-value of 0.90)
These tests suggest that the data support $G_1$-$E$, $G_2$-$E$ and
$G_1$-$G_2$ independence assumption.

%
\begin{sidewaystable}
\textwidth=\textheight
\tablewidth=\textwidth
\caption{}\label{tptable3}
\begin{tabular*}{\tablewidth}{@{\extracolsep{\fill}}@{}p{\tablewidth}@{}}
(a) Analysis results for the MECC study data with statins ($E$), $G_1$
$RS1800775$ on CETP and $G_2$ $RS1056836$ on CYP1B1. The model adjusts
age ($S_1$, $\mbox{``$>$50''}=1$, $\mbox{``$\le$50''}=0$), gender
($S_2$, \mbox{$\mbox{male}=1$}, $\mbox{female}=0$), ethnicity ($S_3$,
$\mbox{Ashkenazi}=1$, $\mbox{Non-Ashkenazi}=0$), sports activity
($S_4$, $\mbox{Yes}=1$, \mbox{$\mbox{No}=0$}), vegetable consumption ($S_5$,
$\mbox{High}=1$, $\mbox{Low}=0$), family history of CRC ($S_6$,
$\mbox{Yes}=1$, $\mbox{No}=0$), the use or nonuse of NSAID within 3
years ($S_7$, $\mbox{Yes}=1$, $\mbox{No}=0$), the use or nonuse of
Aspirin within 3 years ($S_8$, $\mbox{Yes}=1$, $\mbox{No}=0$). Under
the TPFB method the ``est.'' corresponds to the posterior mean, whereas
PSD corresponds to posterior standard deviation. The methods that yield
the smallest PSD are in bold font in each row
\end{tabular*}
\begin{tabular*}{\tablewidth}{@{\extracolsep{\fill}}lk{4.6}k{4.6}k{4.6}
k{4.6}k{4.6}k{4.6}k{4.6}@{}}
\hline
& \multicolumn{1}{c}{\hspace*{-7pt}\textbf{TPFB}}
& \multicolumn{1}{c}{\hspace*{-7pt}\textbf{TPFB}$\bolds{_{\mathrm{emp}}}$}
& \multicolumn{1}{c}{\hspace*{-7pt}\textbf{WL}} & \multicolumn{1}{c}{\hspace*{-7pt}\textbf{PL}}
& \multicolumn{1}{c}{\hspace*{-7pt}\textbf{UML}} & \multicolumn{1}{c}{\hspace*{-7pt}\textbf{CML}}
& \multicolumn{1}{c@{}}{\hspace*{-7pt}\textbf{EB}} 
\\
& \multicolumn{1}{c}{\hspace*{-7pt}\textbf{est.(PSD)}}
& \multicolumn{1}{c}{\hspace*{-7pt}\textbf{est.(PSD)}}
& \multicolumn{1}{c}{\hspace*{-7pt}\textbf{est.(se)}}
& \multicolumn{1}{c}{\hspace*{-7pt}\textbf{est.(se)}}
& \multicolumn{1}{c}{\hspace*{-7pt}\textbf{est.(se)}}
& \multicolumn{1}{c}{\hspace*{-7pt}\textbf{est.(se)}} &
\multicolumn{1}{c@{}}{\hspace*{-7pt}\textbf{est.(se)}} \\
\hline
\multicolumn{5}{@{}l}{Exposure variables} &&
\\
\quad$G_1$ & 0.04,\mbox{ }(0.09) & 0.01,\mbox{ }(0.09) & 0.00,\mbox{ }(0.09)
& 0.00,\mbox{ }(0.09) & 0.00,\mbox{ }(0.09)
& \mbox{\textbf{0.00}},\mbox{ }\mbox{\textbf{(0.08)}} & \mbox{\textbf{$\bolds{-}$0.07}},\mbox{ }\mbox{\textbf{(0.08)}} \\
\quad$G_2$ & -0.04,\mbox{ }(0.10) & -0.06,\mbox{ }(0.10) & -0.13,\mbox{ }(0.10) & -0.13,\mbox{ }(0.10) &
-0.13,\mbox{ }(0.10) & \mbox{\textbf{$\bolds{-}$0.12}},\mbox{ }\mbox{\textbf{(0.08)}} &
\mbox{\textbf{$\bolds{-}$0.12}},\mbox{ }\mbox{\textbf{(0.08)}} \\
\quad Statin use & {-1.29},\mbox{ }(0.30) & \mbox{\textbf{$\bolds{-}$1.32}},\mbox{ }\mbox{\textbf{(0.27)}}
& {-1.30},\mbox{ }(0.30) &
-1.30,\mbox{ }(0.30) & -1.40,\mbox{ }(0.30) & -1.54,\mbox{ }(0.28) & -1.51,\mbox{ }(0.29) \\
\quad$G_1$ x $G_2$ & 0.01,\mbox{ }(0.07) & \mbox{\textbf{0.03}},\mbox{ }\mbox{\textbf{(0.05)}} & 0.05,\mbox{ }(0.08) &
0.06,\mbox{ }(0.08) & 0.06,\mbox{ }(0.08) & 0.06,\mbox{ }(0.06) & 0.06,\mbox{ }(0.06) \\
\quad$G_1$ x statin use &0.34,\mbox{ }({0.17}) & \mbox{\textbf{0.34}},\mbox{ }\mbox{\textbf{({0.15})}}
& 0.25,\mbox{ }(0.18) & 0.25,\mbox{ }(0.18) & 0.25,\mbox{ }(0.18) & \mbox{\textbf{0.38}},\mbox{ }\mbox{\textbf{({0.15})}}
& 0.34,\mbox{ }(0.17) \\
\quad$G_2$ x statin use &\mbox{\textbf{0.33}},\mbox{ }\mbox{\textbf{(0.16)}}
& \mbox{\textbf{0.33}},\mbox{ }\mbox{\textbf{({0.16})}} &
0.38,\mbox{ }(0.18) & 0.38,\mbox{ }(0.19) & 0.38,\mbox{ }(0.20) & 0.38,\mbox{ }(0.18) & 0.38,\mbox{ }(0.19) \\
[4pt]
\multicolumn{5}{@{}l}{Gene-statin and gene--gene association
parameters from $P(G_1, G_2|E, \uS)$}&&\\
\quad$\lambda_{G_1G_2}$ & 0.02,\mbox{ }(0.05) & 0.00,\mbox{ }(0.05) \\
\quad$\lambda_{G_1E}$ & 0.05,\mbox{ }(0.07) & 0.08,\mbox{ }(0.06)\\
\quad$\lambda_{G_2E}$ & 0.01,\mbox{ }(0.07) & 0.01,\mbox{ }(0.07) \\
\hline\\
\end{tabular*}
\begin{tabular*}{\tablewidth}{@{\extracolsep{\fill}}p{\tablewidth}@{}}
(b) Sensitivity analysis with respect to the maximum number of allowable
mixture components $k_{\mathrm{max}}$, and the prior on $G$-$G$ and $G$-$E$
association parameters $\ulambda$
\end{tabular*}
\begin{tabular*}{\tablewidth}{@{\extracolsep{\fill}}llcccccc@{}}
\hline
& & \multicolumn{1}{c}{$\bolds{G_1}$} & \multicolumn{1}{c}{$\bolds{G_2}$}
& \multicolumn{1}{c}{\textbf{Statin use}}
& \multicolumn{1}{c}{$\bolds{G_1}$\textbf{ x }$\bolds{G_2}$}
& \multicolumn{1}{c}{$\bolds{G_1}$\textbf{ x statin use}} &
\multicolumn{1}{c@{}}{$\bolds{G_2}$\textbf{ x statin use}} \\
\hline
$k_{\mathrm{max}}=10$ & TPFB & 0.05 (0.10) & $-$0.03 (0.10) & $-$1.29 (0.30) &
0.01 (0.07) & 0.36 (0.16) & 0.32 (0.17)\\
& TPFB$_{\mathrm{emp}}$ & 0.01 (0.09) & $-$0.06 (0.10) & $-$1.32 (0.29) & 0.03
(0.07) & 0.31 (0.15) & 0.32 (0.16)\\
[4pt]
$k_{\mathrm{max}}=30$ & TPFB$_{\mathrm{non}}$& 0.05 (0.11) & $-$0.03 (0.11) & $-$1.29
(0.31) & 0.01 (0.07) & 0.34 (0.19) & 0.34 (0.21)\\
\hline
\end{tabular*}
\legend{TPFB, TPFB$_{\mathrm{emp}}$, TPFB$_{\mathrm{non}}$:
Two-phase full Bayes [with informative prior $N(0, 10^{-2})$, using
empirical estimates for prior variances, with noninformative prior
$N(0, 10^4)]$ on $G$-$E$ association parameters;
UML: unconstrained maximum likelihood, CML:
constrained maximum likelihood, EB: empirical-Bayes,
WL: weighted likelihood and PL: pseudo-likelihood.}\vspace*{-3pt}
\end{sidewaystable}

We report the results of the multivariate analysis in Table \ref
{tptable3}. Along with the two-phase full Bayes approach (TPFB), we
consider five alternative methods. Unfortunately, none of these
competing methods use the data in both phases and make use of the
independence constraints. The first three use phase II data only (i)
unconstrained maximum likelihood (UML), a retrospective analysis that
does not specify any constraints on $P(G_1,G_2|E, \uS)$, (ii)
constrained maximum likelihood (CML), that imposes the Hardy--Weinberg
equilibrium as well as $G_1$-$E/G_1$-$G_2$ independence, (iii)
empirical-Bayes (EB), using data-adaptive ``shrinkage estimation''
between the constrained and unconstrained ML estimates. Since methods
(ii) and (iii) are developed for single marker analysis, $G_2$-$E$
independence cannot be enforced in existing software [we used the
``CGEN'' package by \citet{BhaChaWhe}].
These three methods completely ignore biased sampling at phase II and
may thus lead to biased estimation of the main effect of $E$,
particularly if the exposure sampling rates were the differential among
cases and controls. The next two approaches use information from both
phases under a prospective likelihood framework: (iv) a
Horvitz--Thompson estimator, typically known as a weighted likelihood
(WL) approach [\citet{ManLer77}, \citet{BreCha99}]. This
approach uses sampling fractions $n_{ij}/N_{ij}$, where $n_{ij}$ and
$N_{ij}$ are the number of subjects corresponding to $D=i, E=j$ at
phases II and I, respectively. The sampling fraction serves as
weights in the likelihood to adjust for biased sampling [we used the
$\textit{svyglm}$ function in the ``survey'' package in R by \citet{Lum}].
Finally, (v) a pseudo-likelihood (PL) approach which also adjusts for
biased sampling probabilities in a likelihood framework
[\citet{Schetal93}]. Briefly, if we denote
$P_{ij}=P(D=i|E=j)=\exp(i\alpha_j)/\{1+\exp(\alpha_j)\}$ where
$\alpha_j$ is the log-odds for $D=1$\vadjust{\goodbreak} when $E=j$, then the
pseudo-likelihood is defined as $\prod_{i, j}P_{ij}^{N_{ij}}
\prod_{i,j,k}p_{ijk}$. Here,
\[
p_{ijk}= \frac{n_{ij}\exp\{i(\beta_0-\alpha_j+s_{ijk}\beta)\}
}{n_{0j}+n_{1j}\exp(\beta_0-\alpha_j+s_{ijk}\beta)},
\]
where $s_{ijk}$ denotes covariate values for a subject with $D=i$ and $E=j$.

Note that all of these five methods use completely observed phase II
data on $G_1$ and $G_2$ as opposed to our proposed method that
includes partially observed data by marginalization of the
likelihood in terms of $G_1$ and $G_2$ when needed.

As previously explained, we present our method (TPFB) corresponding
to two different priors on the $G$-$E$ and $G$-$G$ association
parameters in model (\ref{tploglinear}). First, we consider
informative prior $N(0, 10^{-2})$ that enforces fixed prior belief around
$G$-$E$ and $G$-$G$ independence; we
denote this by TPFB. The analysis using an alternative prior where the
prior variances on $\lambda_{\mathit{GG}}$ and $\lambda_{\mathit{GE}}$ are estimated
based on observed association in the
data is denoted by TPFB$_{\mathrm{emp}}$. In Table~\ref{tptable3}, the
variable selection scheme
is excluded in the TPFB and TPFB$_{\mathrm{emp}}$ by assuming all \mbox{$f_r=1$},
$r=1,\ldots, n_{\beta}$, so that all covariates are included across
all methods. This is done so that the method can be fairly compared to
other alternatives
which do not have the variable selection feature.

Under all methods, note in Table~\ref{tptable3} that the estimated
coefficients corresponding to statin-use suggest strong negative
association with CRC status. The estimated effect size varies
depending on whether the method accounts for biased sampling and/or
gene-environment independence. In the presence of
interactions, we cannot really interpret the main effect estimates
and need to combine the model results to present estimated subgroup
effects. Recall that $G_1$-$E$ and $G_2$-$E$ independence does appear
to be
plausible in light of this data. Note that while $G_2$ x $E$ interaction
is detected by all methods, $G_1$ x $E$ interaction can only be detected
by CML, EB, TPFB and TPFB$_{\mathrm{emp}}$, that is, methods that use the
independence
assumption. The TPFB estimates of terms involving $E$
are slightly different in effect sizes
with smaller standard errors when compared to the other methods.
Smaller standard errors corresponding to interaction
parameters are noted in all retrospective methods that explicitly model
$(G_1,G_2,E)$ dependence structure.

\begin{table}
\tabcolsep=0pt
\caption{Odds ratio estimates (confidence interval
or credible interval)
for CRC corresponding to statin users vs nonusers across genotype
subgroups. Under all five methods,
a model with main effect of $G_1, G_2, E$ controlling for $S$ was fit
as in Table \protect\ref{tptable3}. Common alleles in $G_1$
($RS1800775$ on CETP)
and $G_2$ ($RS1056836$ on CYP1B1) are A and C, respectively, and minor
alleles in $G_1$ and $G_2$ are C and G, respectively}\label{tptable4}
{\fontsize{8.5pt}{11pt}\selectfont{
\begin{tabular*}{\tablewidth}{@{\extracolsep{\fill}}lccccc@{}}
\hline
& \multicolumn{5}{c@{}}{\textbf{Statins}} \\[-4pt]
& \multicolumn{5}{c@{}}{\hrulefill} \\
$\bolds{G_1}$ & \multicolumn{1}{c}{\textbf{A/A}} & \multicolumn{1}{c}{\textbf{A/C}}
& \multicolumn{1}{c}{\textbf{C/C}} & \multicolumn{1}{c}{\textbf{A/A}}
& \multicolumn{1}{c@{}}{\textbf{A/A}}\\
$\bolds{G_2}$ & \multicolumn{1}{c}{\textbf{C/C}} & \multicolumn{1}{c}{\textbf{C/C}}
& \multicolumn{1}{c}{\textbf{C/C}} & \multicolumn{1}{c}{\textbf{G/C}} &
\multicolumn{1}{c@{}}{\textbf{G/G}}\\
\hline
TPFB & 0.27 (0.15, 0.39) & 0.35 (0.22, 0.44) & 0.48 (0.26, 0.65) & 0.38
(0.24, 0.51) & 0.48 (0.30, 0.77) \\
TPFB$_{\mathrm{emp}}$ & 0.26 (0.15, 0.42) & 0.36 (0.23, 0.45) & 0.49 (0.31,
0.69) & 0.37 (0.23, 0.50) & 0.50 (0.31, 0.79) \\
WL & 0.27 (0.15, 0.49) & 0.35 (0.22, 0.55) & 0.45 (0.26, 0.77) & 0.40
(0.26, 0.62) & 0.59 (0.34, 1.02) \\
PL & 0.27 (0.15, 0.49) & 0.35 (0.22, 0.55) & 0.45 (0.26, 0.79) & 0.40
(0.25, 0.63) & 0.59 (0.33, 1.05) \\
[4pt]
UML & 0.25 (0.14, 0.44) & 0.32 (0.20, 0.50) & 0.41 (0.23, 0.72) & 0.36
(0.23, 0.57) & 0.53 (0.29, 0.95) \\
CML & 0.22 (0.12, 0.37) & 0.33 (0.21, 0.51) & 0.49 (0.29, 0.83) & 0.34
(0.20, 0.48) & 0.46 (0.26, 0.80) \\
EB & 0.22 (0.13, 0.39) & 0.31 (0.20, 0.49) & 0.43 (0.24, 0.79) & 0.32
(0.21, 0.49) & 0.47 (0.27, 0.82) \\
\hline
\end{tabular*}}}
\legend{TPFB, TPFB$_{\mathrm{emp}}$: Two-phase full Bayes
(with empirical estimates for prior variances), UML: unconstrained
maximum likelihood, CML: constrained maximum
likelihood, EB: empirical-Bayes, WL: weighted likelihood,
and PL: pseudo-likelihood.}
\end{table}

We also carried out a sensitivity analyses with respect to the choice
of threshold to truncate the maximum
value of $k$ in the DX construction and the prior on $G$-$E$ and
$G$-$G$ association. As can be seen from
Table~\ref{tptable3}(b), the results are almost identical with
a smaller number ($k_{\mathrm{max}}=10$) of components in the mixture
distribution for $W$. This suggests
further computational efficiency gain is possible by imposing more
parsimonious constraint on $k$. In another sensitivity analysis, when
the prior
on $G$-$E$ association is noninformative $N(0,10^4)$, we
notice TPFB estimates slightly drift toward the estimates from PL
and WL while losing some efficiency on the $G_1$ x $E$ and $G_2$ x $E$ terms.\vadjust{\goodbreak}

\begin{figure}

\includegraphics{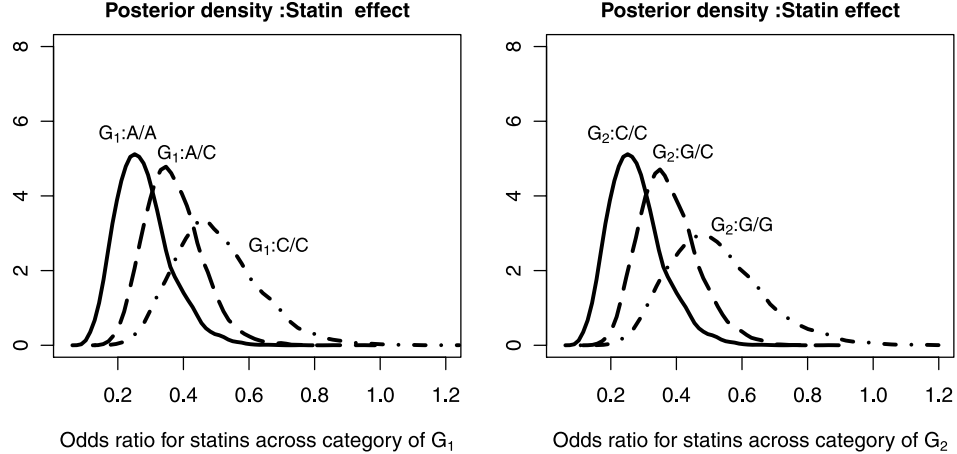}

\caption{The left figure shows the posterior densities of the odds
ratio estimates of CRC corresponding to statin users versus nonusers
across three genotypes in $RS1800775$ on CETP$(G_1)$, holding the
genotype in $RS1056836$ on CYP1B1 at the most frequent category, that
is, $(G_2)=(\mbox{C/C})$. Similarly, the right figure shows the posterior
densities of the odds ratio estimates corresponding to statin users
versus nonusers across three genotypes in $RS1056836$ of CYP1B1$(G_2)$,
holding the genotype in $RS1800775$ of CETP fixed at the most frequent
category, that is, $(G_1)=(\mbox{A/A})$.}\label{tpfigure4}
\end{figure}

To reflect our main interest in subgroup effects of statin across
genotype configurations, we report effects of statin across genotype
subgroups of one SNP, holding the other SNP fixed at the common
genotype category for that second SNP (coded as 0) in Table
\ref{tptable4}. It seems that statin effect is
strongly modified by $G_1$ and $G_2$. According to
TPFB$_{\mathrm{emp}}$ estimates, keeping the $G_2$ genotype fixed at
C/C, the benefit of taking statins to reduce the risk of CRC is maximum
in the A/A genotype of $G_1$ with the posterior estimate (and 95\% HPD)
of the odds-ratios (corresponding to statin users versus nonusers)
being 0.26 (0.15, 0.42). The corresponding ORs in genotype category A/C
and C/C are 0.36 (0.23, 0.45) and 0.49 (0.31, 0.69), respectively.
Figure~\ref{tpfigure4} illustrates estimated posterior densities of the
odds ratios corresponding to statin-use across each genotype of $G_1$
(left) or $G_2$ (right), respectively, while holding the other SNP
fixed at the most common category. This figure indicates that the
protective effect of statin in CRC is diminishing as the allelic dosage
for the minor allele increases in both $G_1$ and $G_2$. Overall, the
TPFB approaches provide much narrower credible intervals compared to PL
and WL by exploiting $G_1$-$E$ and $G_2$-$E$ independence. The
estimates from methods that use phase II data only, like CML, UML and
EB, are numerically slightly different.\vspace*{6pt}

\textsc{Variable selection}: We explore how
variable selection feature performs in this example for the TPFB method.
Previous research by \citet{IshRao03} discussed the
performance of spike and slab prior for general variable selection. We
introduce three
SNPs (RS5925224, RS10174721, RS10077453) and all possible pairwise
$G$ x $G$ and $G$ x $E$ interactions to the previous two SNP model
as fit in Table~\ref{tptable3}. The dimension of the disease risk
model is now 34.
None of the main effects and interactions corresponding to these
three additional SNPs were found significant in an initial single
marker analysis.

We set $f_r =1$ for $S_1$ through $S_8$ to always keep the
confounders and precision variables in the model. The tuning
parameters $v_0$ are fixed at $0.0001$ for this application with sensitivity
analysis results presented for $v_0=0.001$ in Table~\ref{tptable6}.
We would like to see if the variable selection can still detect a
significant $G_1$ x $E$ and $G_2$ x $E$ interaction. Moreover, we would
like to assess
if the three additional SNPs and the corresponding interactions we
added (with null effects as observed
in our initial analysis) are also identified to be not informative by
this process. We tabulate the posterior
distribution of $f_r=1$ among $\uf=(f_1,\ldots, f_{n_\beta})$ which indicate
``in-and-out'' frequencies of the corresponding parameters. These
posterior frequencies of $f_r=1$ can be used to define a ranking of
important predictors. An alternative is to rank the top models (not
just the predictors individually). Before implementing the TPFB, we
reduced the dimensionality of parameters in the model $P(\uG|\uW)$
where $\uG=(G_1, G_2, G_3, G_4, G_5)$ by assuming common
$\lambda_{\mathit{GG}}$ and $\lambda_{\mathit{GE}}$ association parameters across all
SNPs. We use $N(0, 0.1^2)$ prior on this common parameter. In addition,
we further assume a single
common parameter $\lambda_{\mathit{GS}}$ for all $G$-$S$ associations with a vague
normal prior $N(0, 10^4)$. These are assumptions that may be stringent
in certain
situations, but to reduce estimation burden in the log-linear model, we
do need
to make these assumptions for the TPFB methods. For SNPs on a same
functional pathway like in our example,
it may not be too unrealistic to assume a shared association parameter
across SNPs.

\begin{table}
\caption{}\label{tptable6}
\begin{tabular*}{\tablewidth}{@{\extracolsep{\fill}}@{}p{\tablewidth}@{}}
(a) The top 10 promising models in terms of
estimated posterior probabilities of the models. All $\uS$ adjustment
variables are retained in the
model by default and variable selection is performed only on the five
genetic and environmental factors and all
possible pairwise interactions.
Bayesian Information Criterion (BIC) is provided for each model.
Results in parentheses represent the
sensitivity analysis carried out with $v_0=0.001$
\end{tabular*}
\begin{tabular*}{\tablewidth}{@{\extracolsep{\fill}}lcc@{}}
\hline
\textbf{Model} & \multicolumn{1}{c}{\textbf{Posterior probability \%}} & \textbf{BIC} \\
\hline
{[$E$]}[All $S$][$G_2$ x $E$] & 13.1\% (12.5\%)& 43,992 (43,967) \\
{[$E$]}[All $S$][$G_1$ x $E$][$G_2$ x $E$] & 9.2\% (6.3\%)& 43,993
(43,978)\\
{[$E$]}[All $S$][$G_1$ x $E$] & 7.7\% (5.1\%) & 43,994 (43,967) \\
{[$E$]}[All $S$] & \hphantom{0}7.5\% (10.6\%)& 43,997 (43,977)\\
{[$E$]}[All $S$][$G_2$ x $E$][$G_3$ x $E$] & 3.9\% (4.6\%) & 44,004
(43,977) \\
{[$E$]}[All $S$][$G_2$ x $E$][$G_4$ x $E$] & 2.4\% (2.2\%) & 44,002
(43,974)\\
{[$E$]}[All $S$][$G_2$ x $E$][$G_3$ x $G_4$] & 2.1\% (1.5\%) & 43,998
(43,971)\\
{[$E$]}[All $S$][$G_1$ x $E$][$G_3$ x $E$] & 2.1\% (2.2\%) & 44,005
(43,974)\\
{[$E$]}[All $S$][$G_1$][$G_2$ x $E$] & 1.8\% (0.7\%) & 43,996
(43,976)\\
{[$E$]}[All $S$][$G_1$ x $E$][$G_2$ x $E$][$G_5$ x $E$] & 1.6\% (2.0\%)
& 44,010 (43,974)\\
\hline
\end{tabular*}
\legend{BIC represents Bayesian Information Criterion.}\vspace*{10pt}
\begin{tabular*}{\tablewidth}{@{\extracolsep{\fill}}p{\tablewidth}@{}}
(b) The estimated posterior probabilities of appearance
corresponding to $\uG$ and $\uE$ main effects and their interactions are
shown under the identical setting as in Table~\ref{tptable6}(a).
Results in parentheses represent the
sensitivity analysis carried out with $v_0=0.001$
\end{tabular*}
\tabcolsep=0pt
\begin{tabular*}{\tablewidth}{@{\extracolsep{\fill}}lcccccccccc@{}}
\hline
$\bolds{G_1}$ & $\bolds{G_2}$ & $\bolds{G_3}$ & $\bolds{G_4}$
& $\bolds{G_5}$ & $\bolds{E}$ & $\bolds{E}$ \textbf{x} $\bolds{G_1}$
& $\bolds{E}$ \textbf{x} $\bolds{G_2}$
& $\bolds{E}$ \textbf{x} $\bolds{G_3}$ &
$\bolds{E}$ \textbf{x} $\bolds{G_4}$ & $\bolds{E}$ \textbf{x} $\bolds{G_5}$\\
\hline
\hphantom{(}6.7 & 5.0 & \hphantom{0}4.3 & 4.1 & 7.6 & 100.0 & 36.8 & 64.0 & 18.4 & 10.1 & 13.9\\
(5.6) & (6.8) & (10.6) & (7.3) & (9.4) & 100.0 & (29.5) & (55.4) &
(19.9) & \hphantom{0}(9.7) & \hphantom{0}(9.5)\\
\hline
\end{tabular*}
\end{table}
%

In Table~\ref{tptable6}, we present numerical results on model and
predictor ranking as well as the Bayesian Information Criterion
(BIC) corresponding to each model. We only present the top 10
models. According to the result, the model with main effects of $E$
and $G_2$ x $E$ interactions seems to be the
most preferred model (posterior probability 13.1\%) followed by the model
with E and both $G_1$ x $E$ and $G_2$ x $E$ interactions (posterior probability
9.2\%). With $v_0=0.001$, the ranking of predictors is slightly
different, as the main effects of $G_1$ through $G_5$ are now selected
more often.
The bottom panel of Table~\ref{tptable6} shows the frequency of
retaining a predictor in the model according to the posterior
distribution of $\uf$. The main effect of $E$ appears most of the
times (100\%) with large selection probabilities for $G_1$ x $E$
and $G_2$ x $E$ interactions (36.8\% and 64.0\%), respectively.
Overall, nonsignificant interactions/main effects are well filtered
under this variable selection scheme.

\section{Simulation study}\label{tpsim}
In this section we assess the performance of the proposed method by
conducting a simulation study. We mainly consider two aspects: (i)
varying gene-gene/gene-environment association structure and (ii)
when phase~II sampling is the differential between cases and controls.
We compare our method with the
five alternative methods mentioned before: WL, PL, UML, CML and EB
in terms of the average bias and mean squared errors (MSE), based on
1000 simulated data sets.

We first describe the data generation procedure. We consider two
genes $G_1$ and~$G_2$, and one environment factor $E$, with disease
status $D$, all binary. We generate data from the following
log-linear model [\citet{LiCon09}]:
%
\begin{eqnarray}
\label{tpsimlin}\qquad \log(\mu|D, G_1, G_2, E) &=&
\gamma_0+ \gamma_{G_1}G_1+
\gamma_{G_2}G_2 + \gamma_{E}E +
\gamma_{D}D
\nonumber\\
&&{}+ \lambda_{G_1 E}G_1 E + \lambda_{G_2 E}G_2
E + \lambda_{G_1 G_2} G_1 G_2
\nonumber\\[-8pt]\\[-8pt]
&&{}+ \beta_{G_1}G_1 D + \beta_{G_2}G_2
D + \beta_{E}E D
\nonumber
\\
&&{}+ \beta_{G_1E}G_1 E D + \beta_{G_2E}G_2
E D + \beta_{G_1G_2} G_1 G_2 D,
\nonumber
\end{eqnarray}
where $\mu$ denotes expected cell counts corresponding to the $(D,
G_1, G_2, E)$ configuration. Under this model, we are capable of
fixing $G_1$-$E$, $G_2$-$E$ and $G_1$-$G_2$ association under
controls by setting values of $\lambda_{G_1E}, \lambda_{G_2E}$ and
$\lambda_{G_1G_2}$, respectively. These parameters are
approximately equivalent to those in model $P(G_1, G_2|W)$
(\ref{tploglinear}) when the disease is rare. Similarly, we can set
$\beta_{G_1E}, \beta_{G_2E}$ or $\beta_{G_1G_2}$, corresponding to the
$G$ x $E$ or $G$ x $G$ interactions in the disease risk model.
The parameters $(\gamma_0,\gamma_{G_1},\gamma_{G_2}, \gamma_{E})$
control the marginal frequencies of $G_1$, $G_2$ and $E$ in controls. A large
negative value of $\gamma_D$ ensures that the
disease is rare.

For the model parameters in (\ref{tpsimlin}), we fixed $(\gamma_0,
\gamma_{G_1}, \gamma_{G_2}, \gamma_{E}, \gamma_{D}) =(-6,\break -0.5,
-0.5,-2.0, -4.5)$ that produces approximately 2.5\% of the cases,
frequency of $G_1=1$ and $G_2=1$ both at 45\% while the prevalence
of $E=1$ is 15\%. We assign $(\beta_{G_1E}, \beta_{G_2E}, $
$\beta_{G_1G_2}) =(0, \log(2), \log(2))$ in (\ref{tpsimlin}). For
setting parameters corresponding to $G$-$E/G$-$G$ association, we
set $(\lambda_{G_1G_2},\lambda_{G_1E},\break \lambda_{G_2E} ) =(\log(2),
0$, $\log(1.5))$ to reflect $G_1$-$G_2$ and $G_2$-$E$
dependence, and $(0, 0, 0)$ for the independence scenario.

Now we turn our attention to the sampling design. We randomly
generate $1000$ cases and $1000$ controls with complete $(D, G_1,
G_2, E)$ data. We then carry out $(D, E)$-stratified sampling as
follows. We select 600 cases and 600 controls in
phase II. We consider two scenarios regarding this
the stratified sampling strategy: (a) all subjects with a positive
$E(\mbox{$=$}1)$, in cases
and controls, are automatically included in phase II; (b)
all subjects with a positive $E(\mbox{$=$}1)$ in cases are included in phase II,
however, 600 controls for phase II are randomly selected regardless of $E$
status. Finally, information on $G_1$ and $G_2$ from phase I subjects,
that is, 400 cases and 400 controls, is treated as missing by design.
We iterate
this step to generate 1000 replicate data sets under each sampling scheme.

%
\begin{sidewaystable}
\tabcolsep=3pt
\textwidth=\textheight
\tablewidth=\textwidth
\caption{Simulation results under exposure
enriched sampling
with all $E=1$ in phase I data selected in phase II for both cases and controls.
We consider two association scenarios: (1)
$G_1 \bott E$, $G_1 \bott G_2$ and $G_2 \bott E$ association, (2) $G_1
\bott E$, $G_1 \sim G_2$ and $G_2 \sim E$. The results are based on
1000 replicated data sets, each with 1000 cases and 1000 controls in
phase I and 600 cases and 600 controls in phase II.
The approaches listed, TPFB, TPFB$_{\mathrm{emp}}$, WL, PL, UML,
CML and EB, each represent two-phase full Bayes (with
empirically obtained prior variance), weighted likelihood, pseudo-likeliohod,
unconstrained maximum likelihood, constrained maximum likelihood,
and empirical-Bayes, respectively. The CML imposes $G_1$-$E$ and
$G_1$-$G_2$ independence, however, no constraints on $G_2$-$E$
association. We set $(\beta_{E}, \beta_{G_1G_2}, \beta_{G_1E}, \beta_{G_2E})=
(-1.5, 0, \log(2), \log(2))$ for all scenarios. The rows with the
smallest two sum (MSE)
are in bold}\label{tptable5a}
\begin{tabular*}{\tablewidth}{@{\extracolsep{\fill}}lcd{2.4}d{2.4}d{2.4}
d{2.4}cd{2.4}cd{2.4}cc@{}}
\hline
& & \multicolumn{5}{c}{$\bolds{G_1 \bott E}$, $\bolds{G_1 \bott G_2}$,
$\bolds{G_2 \bott E}$} &
\multicolumn{5}{c@{}}{$\bolds{G_1 \bott E}$, $\bolds{G_1 \sim G_2}$, $\bolds{G_2 \sim E}$}
\\[-4pt]
& & \multicolumn{5}{l}{\hspace*{-2pt}\hrulefill} & \multicolumn{5}{l@{}}{\hspace*{-2pt}\hrulefill} \\
\multicolumn{2}{@{}l}{\textbf{Stratified sampling (a)}\tabnoteref{ta}}
& \multicolumn{1}{c}{\hspace*{-2pt}$\bolds{E}$} &
\multicolumn{1}{c}{\hspace*{-3pt}$\bolds{G_1}$ \textbf{x}
$\bolds{G_2}$} & \multicolumn{1}{c}{\hspace*{-3pt}$\bolds{G_1}$ \textbf{x} $\bolds{E}$}
& \multicolumn{1}{c}{\hspace*{-3pt}$\bolds{G_2}$ \textbf{x} $\bolds{E}$}
& \multicolumn{1}{c}{\textbf{Sum (MSE)\tabnoteref{tb}}} &
\multicolumn{1}{c}{\hspace*{-2pt}$\bolds{E}$} &
\multicolumn{1}{c}{$\bolds{G_1}$ \textbf{x} $\bolds{G_2}$}
& \multicolumn{1}{c}{\hspace*{-3pt}$\bolds{G_1}$ \textbf{x} $\bolds{E}$}
& \multicolumn{1}{c}{$\bolds{G_2}$ \textbf{x} $\bolds{E}$}
& \multicolumn{1}{c@{}}{\textbf{Sum (MSE)\tabnoteref{tb}}} \\
\hline
& & \multicolumn{5}{c}{$(\lambda_{G_1G_2}, \lambda_{G_1E}, \lambda_{G_2E})=(0,0,0)$}
& \multicolumn{5}{c@{}}{$(\lambda_{G_1G_2}, \lambda_{G_1E}, \lambda_{G_2E})=(\log(2),0,\log(1.5))$}
\\[4pt]
TPFB & Bias & -0.024 & -0.017 & 0.020 & -0.017 && -0.056 & 0.166 &
-0.022 & 0.119 & \\
& (MSE) & (0.093) & (0.044) & (0.120) & (0.135) &(0.392)&(0.117) &
(0.081) & (0.091) & (0.122) & (0.411)\\
TPFB$_{\mathrm{emp}}$ & Bias & 0.007 & -0.019 & -0.021 & -0.062 && -0.033 &
0.043 & -0.029 & 0.026 & \\
& (MSE) & (0.089) & (0.025) & (0.111) & (0.126) &(\textbf{0.351})
&(0.113) & (0.064) & (0.091) & (0.120) &(\textbf{0.388}) \\
WL & Bias & -0.038 & -0.025 & 0.043 & 0.009 && -0.038 & 0.011 & 0.011
& 0.006 & \\
& (MSE) & (0.099) & (0.058) & (0.144) & (0.157) &(0.458)&(0.105) &
(0.057) & (0.101) & (0.121) &(0.384) \\
PL & Bias & -0.038 & -0.026 & 0.043 & 0.009 && -0.038 & 0.011 & 0.011
& 0.006 & \\
& (MSE) & (0.098) & (0.056) & (0.144) & (0.157) &(0.455)&(0.105) &
(0.056) & (0.101) & (0.121) & (\textbf{0.383})\\
UML & Bias & -0.093 & -0.026 & 0.043 & 0.009 && -0.096 & 0.011 & 0.011
& 0.006 & \\
& (MSE) & (0.110) & (0.056) & (0.144) & (0.157) &(0.467)& (0.116) &
(0.056) & (0.101) & (0.121) &(0.394) \\
CML & Bias & -0.085 & -0.020 & 0.026 & 0.003 && -0.100 & 0.700 & 0.011
& 0.009 & \\
& (MSE) & (0.099) & (0.025) & (0.083) & (0.155) &(\textbf
{0.362})&(0.112) & (0.520) & (0.070) & (0.116) & (0.818)\\
EB & Bias & -0.087 & -0.025 & 0.036 & 0.004 && -0.099 & 0.089 & 0.010
& 0.008 & \\
& (MSE) & (0.099) & (0.036) & (0.099) & (0.155) &(0.389)&(0.112) &
(0.069) & (0.075) & (0.116) &(0.392) \\
\hline
\end{tabular*}
\tabnotetext[\mbox{$\dag$}]{ta}{All subjects with $E=1$ in case and control
are subsampled for phase II.}
\tabnotetext[\mbox{$\ast$}]{tb}{The combined MSEs as summed over all four
parameters.}
\legend{TPFB uses the informative prior $N(0, 10^{-2})$ on
$G$-$G$ and $G$-$E$ associations in the model (\ref{tploglinear}).
TPFB$_{\mathrm{emp}}$ uses the prior $N(0, {\hat
{\theta}^{2}})$ on $G$-$G$ and $G$-$E$ associations in the model (\ref
{tploglinear}) where
${\hat{\theta}^{2}}$ is empirically estimated as
the $G$-$G$ or $G$-$E$ association parameter under the controls.}
\end{sidewaystable}
%

%
\begin{sidewaystable}
\tabcolsep=3pt
\textwidth=\textheight
\tablewidth=\textwidth
\caption{Simulation results under exposure enriched sampling with all
$E=1$ in phase I data selected in phase II for cases but\break a random
sample of controls are selected in phase II. We consider two
association scenarios: (1) $G_1 \bott E$, $G_1 \bott G_2$,\break and $G_2 \bott
E$ association, (2) $G_1 \bott E$, $G_1 \sim G_2$, and $G_2 \sim E$. The
results are based on 1000 replicated data sets, each with\break 1000 cases
and 1000 controls in phase I and 600 cases and 600 controls in phase
II. The approaches listed, TPFB,\break TPFB$_{\mathrm{emp}}$, WL, PL, UML,
CML, and EB, each represent two-phase full Bayes (with empirically
obtained prior variance),\break weighted likelihood, pseudo-likeliohod,
unconstrained maximum likelihood, constrained maximum likelihood, and
empirical-Bayes, respectively. The CML imposes $G_1$-$E$ and
$G_1$-$G_2$ independence, however, no constraints on the $G_2$-$E$
association. We set $(\beta_{E}, \beta_{G_1G_2}, \beta_{G_1E},
\beta_{G_2E})= (-1.5, 0, \log(2), \log(2))$ for all scenarios. The rows
with the smallest two sum (MSE) are in bold}\label{tptable5b}
\begin{tabular*}{\tablewidth}{@{\extracolsep{4in minus 4in}}lcd{2.4}cd{2.4}
d{2.4}cd{2.4}ccd{2.4}c@{}}
\hline
& & \multicolumn{5}{c}{$\bolds{G_1 \bott E}$, $\bolds{G_1 \bott G_2}$,
$\bolds{G_2 \bott E}$} &
\multicolumn{5}{c@{}}{$\bolds{G_1 \bott E}$, $\bolds{G_1 \sim G_2}$, $\bolds{G_2 \sim E}$}
\\[-4pt]
& & \multicolumn{5}{l}{\hspace*{-2pt}\hrulefill}
& \multicolumn{5}{l@{}}{\hspace*{-2pt}\hrulefill} \\
\multicolumn{2}{@{}l}{\textbf{Stratified sampling (b)\tabnoteref{tc}}}
& \multicolumn{1}{c}{\hspace*{-2pt}$\bolds{E}$} &
\multicolumn{1}{c}{$\bolds{G_1}$ \textbf{x}
$\bolds{G_2}$} & \multicolumn{1}{c}{\hspace*{-3pt}$\bolds{G_1}$ \textbf{x} $\bolds{E}$}
& \multicolumn{1}{c}{\hspace*{-3pt}$\bolds{G_2}$ \textbf{x} $\bolds{E}$}
& \multicolumn{1}{c}{\textbf{Sum (MSE)\tabnoteref{td}}} &
\multicolumn{1}{c}{\hspace*{-2pt}$\bolds{E}$} &
\multicolumn{1}{c}{$\bolds{G_1}$ \textbf{x} $\bolds{G_2}$}
& \multicolumn{1}{c}{$\bolds{G_1}$ \textbf{x} $\bolds{E}$}
& \multicolumn{1}{c}{\hspace*{-3pt}$\bolds{G_2}$ \textbf{x} $\bolds{E}$}
& \multicolumn{1}{c@{}}{\textbf{Sum (MSE)\tabnoteref{td}}} \\
\hline
& & \multicolumn{5}{c}{$(\lambda_{G_1G_2}, \lambda_{G_1E}, \lambda_{G_2E})=(0,0,0)$} &
\multicolumn{5}{c@{}}{$(\lambda_{G_1G_2}, \lambda_{G_1E},
\lambda_{G_2E})=(\log(2),0,\log(1.5))$} \\
[4pt]
TPFB & Bias & 0.007 & 0.022 & -0.007 & -0.022 && -0.105 & 0.160 &
0.032 & 0.128 & \\
& (MSE) & (0.081) & (0.040) & (0.113) & (0.124) &(\textbf{0.358})&
(0.125) & (0.073) & (0.106) & (0.128) &(\textbf{0.432}) \\
TPFB$_{\mathrm{emp}}$ & Bias & 0.039 & 0.015 & -0.054 & -0.073 && -0.027 &
0.036 & 0.024 & 0.000 & \\
& (MSE) & (0.086) & (0.031) & (0.127) & (0.122) &(\textbf{0.366})
&(0.121) & (0.058) & (0.115) & (0.135) & (\textbf{0.429})\\
WL & Bias & -0.012 & 0.016 & 0.021 & 0.024 && -0.044 & 0.004 & 0.076 &
-0.014 & \\
& (MSE) & (0.098) & (0.059) & (0.165) & (0.167) &(0.489)&(0.133) &
(0.056) & (0.150) & (0.147) & (0.486)\\
PL & Bias & -0.012 & 0.015 & 0.020 & 0.025 && -0.046 & 0.002 & 0.077 &
-0.013 & \\
& (MSE) & (0.097) & (0.059) & (0.164) & (0.166) &(0.486)&(0.132) &
(0.055) & (0.149) & (0.146) & (0.482)\\
UML & Bias & 0.538 & 0.015 & 0.020 & 0.025 && 0.520 & 0.002 & 0.077 &
-0.013 & \\
& (MSE) & (0.395) & (0.059) & (0.164) & (0.166) &(0.784)&(0.407) &
(0.055) & (0.149) & (0.146) &(0.757) \\
CML & Bias & 0.544 & 0.017 & -0.002 & 0.018 && 0.530 & 0.699 & 0.046 &
-0.013 & \\
& (MSE) & (0.384) & (0.030) & (0.088) & (0.161) &(0.663)&(0.399) &
(0.515) & (0.073) & (0.141) &(1.128) \\
EB & Bias & 0.543 & 0.016 & 0.008 & 0.018 && 0.528 & 0.078 & 0.059 &
-0.013 & \\
& (MSE) & (0.385) & (0.039) & (0.112) & (0.161) &(0.697)&(0.401) &
(0.066) & (0.097) & (0.141) & (0.705) \\
\hline
\end{tabular*}
\tabnotetext[\mbox{$\dag$}]{tc}{All cases with $E=1$ are included in phase
II, however, controls are randomly selected for phase II.}
\tabnotetext[\mbox{$\ast$}]{td}{The combined MSEs over all four parameters.}
\legend{TPFB uses the informative prior $N(0, 10^{-2})$ on
the $G$-$G$ and $G$-$E$ associations in the model (\ref{tploglinear}).
TPFB$_{\mathrm{emp}}$ uses the prior $N(0, {\hat
{\theta}^{2}})$ on $G$-$G$ and $G$-$E$ associations in the model
(\ref{tploglinear}) where ${\hat{\theta}^{2}}$ is empirically estimated as
the $G$-$G$ or $G$-$E$ association parameter under the controls.}
\end{sidewaystable}

Tables~\ref{tptable5a} and~\ref{tptable5b} display the simulation
results based on two different sampling schemes (a) and (b), respectively.
We follow the convention that $\bot$ and $\sim$ represent
independence and
dependence between two variables, respectively. Under $G_1 \bott E, G_2
\bott E$ and $G_1
\bott G_2$ the CML method yields the smallest MSE with respect to
$G_1$ x $E$ and $G_1$ x $G_2$ interaction followed by TPFB,
TPFB$_{\mathrm{emp}}$ and $EB$, while WL, PL and UML present relatively
larger MSE. Here we need to note that the current implementation of CML
and EB can only use $G_1$-$E$ and $G_1$-$G_2$ independence, but not $G_2$-$E$
independence. As phase II sampling becomes differential between
cases and controls from scenario (a) (Table~\ref{tptable5a}) to (b)
(Table~\ref{tptable5b}), we notice the substantial increase
in the bias for estimating the main effect of $E$ from CML, UML and EB
as expected, while
WL, PL, TPFB and TPFB$_{\mathrm{emp}}$ provide relatively less biased
estimates. This trend remains present in the case where $G_1 \bott E, G_2
\sim E$ and $G_1 \sim G_2$. Beyond the bias in $\beta_E$ from
CML, UML and EB, we note that under the departure from the
independence assumption, namely, $G_1 \bott G_2$, there is a dramatic
increase in the bias corresponding to the $G_1$ x $G_2$ interaction
under CML and to some extent in TPFB. TPFB$_{\mathrm{emp}}$ and EB are more
robust to this assumption. Both TPFB show gain in efficiency for
interaction estimation compared to PL and~WL. Overall, our proposed
methods, especially TPFB$_{\mathrm{emp}}$, yield obvious gain in efficiency
compared to PL and WL in terms of the $G$ x $E$ or $G$ x $G$
interactions in the presence of independence. On the other hand,
TPFB$_{\mathrm{emp}}$ provides less biased estimates of the $E$ effect
compared to
UML, CML and EB which use only phase II data. When the subsampling
ratio is 80\%, the pattern
remains the same as seen in online supplemental Table 2 [\citet{Ahnetal}].
We also provide the sum of the MSEs across all parameters in order to capture
the accuracy of estimating subgroup effects defined by different $G$-$E$
configurations. This summary measure in the last columns of Tables~\ref{tptable5a} and
\ref{tptable5b}
clearly suggests that our methods yield more efficient characterization
of the
joint effect of exposure and genetic factors.

Table 3 in the supplemental article [\citet{Ahnetal}] presents
simulation results
under the traditional or unstratified case-control design when
a random sample of cases and controls are taken irrespective
of $E$ status. We can note clear efficiency gains from stratified
sampling when comparing Table 4 to Table 3 in the supplemental article
for estimating the interaction parameters.

\section{Discussion}\label{tpdiscuss}
We presented a flexible Bayesian approach to estimate gene--gene ($G$
x $G$) and/or gene-environment ($G$ x $E$) interactions under
two-phase sampling with multiple markers. The proposed approach can
handle multiple
genetic and environmental factors. The method can trade off between
bias and efficiency by incorporating uncertainty around
gene-environment independence through the hierarchical structure in
a data-adaptive way. The underlying ingredients of this hierarchy
are the disease risk model, the multivariate gene model and the
joint model for the environment factors/covariates, respectively. Our
method can also handle potential missingness in genetic information
due to technical inconsistency or due to merging different studies
or cohorts, leading to nonmonotone missing data structure at the phase
II subsample. This
paper is the first Bayesian paper with retrospective modeling for
$G$ x $E$ studies under two-phase sampling that can handle multiple
markers.

We compared our method to simpler alternatives such as UML, CML and
EB that use gene-environment independence but only based on phase
II data, ignoring biased sampling. We also considered methods that
account for biased sampling at phase II: weighted likelihood and
pseudo-likelihood, but do not leverage the independence assumption.
Our method provides a framework that integrates both of these
features. In a clinical study like the MECC example, where interest
lies in estimating the differential effect of statin use across
genetic subgroups for devising targeted prevention strategies,
estimates of main effects as well as gene-environment interaction
are equally important, thus both estimates need to be assessed. In terms
of aggregate MSE, our method has superior performance across a wide
range of scenarios over the competing method.

There are some limitations of the current paper that need to be
expanded and explored in future studies. First, we do not fully
address the performance of our method in the presence of a truly
high-dimensional gene model through simulation studies. The method
is scalable to handle up to 294 SNPs and pairwise interactions in
our data example, but we have not carried out a simulation study due
to computation time. We also need to deal with the exponentially
increasing number of $G$ x $E$ and $G$ x $G$ interactions in the
disease risk model as well as $G$-$E$/$G$-$G$/$G$-$S$ associations
in the multivariate gene model, as we add more $G$-variables in the
model. We address this by Bayesian variable selection and assuming a
common parameter for $G$-$E$/$G$-$G$/$G$-$S$ association on genes in
the same pathway in the multivariate gene model. The latter is
a rather ad-hoc strategy for reducing the dimension and is a limitation of
our method. Bias in parameter
estimates is expected to arise under departures from this
assumption. Calculation of $P(D)$ in the denominator of the likelihood
could also pose challenges with truly high-dimensional data. Second, we
have not tested the
\citet{BhaDun} algorithm for the mixed set of discrete
and continuous covariates in~$W$. Future research will focus on the
higher-dimensional $G$ and $E$ settings, more general structure of
the $W$ vector as well as the possibility of capturing higher order
interactions, not just pairwise interactions.

For practitioners who want to choose a design strategy to enhance the
power of screening $G$ x $E$ effects with a relatively rare exposure,
exposure enrichment of cases and controls for collecting genotype data
is a better strategy than random sampling. The tools we developed in
the paper provides a way to account for the biased sampling. The
approach also allows one to explore a multivariate model with multiple
SNPs and environmental exposure and identify potentially informative
predictors. If the interest lies in characterizing subgroup effects of
$E$ across different subgroups defined by $\uG$, this design and
analysis strategy is particularly powerful. We recommend the use of
default prior choices in the codes available at
\url{http://www.umich.edu/\textasciitilde jaeil/tp.zip} and recommend
using TPFB$_{\mathrm{emp}}$ as the analysis to be reported. For
prescribing a preventive medicine prophylactically, like use of statins
for colorectal cancer, identifying genetic subgroups that will receive
the most benefit from such a therapy is particularly helpful.
Characterizing $G$ x $E$ effects furthers our understanding of such
subgroup effects for tailoring targeted prevention strategies.

\begin{supplement}
\stitle{Bayesian semiparametric analysis for two-phase studies of
gene-environ\-ment
interaction}
\slink[doi]{10.1214/12-AOAS599SUPP} 
\sdatatype{.pdf}
\sfilename{aoas599\_supp.pdf}
\sdescription{We consider two-phase studies of $G$ x $E$ interaction where
phase I data is available on exposure,
covariates and disease status and stratified sampling is done to
prioritize individuals for genotyping at phase II.
We consider a Bayesian analysis based on the joint retrospective
likelihood of phases I and II data
that handles multiple genetic and environmental factors, data adaptive
use of gene-environment independence.}
\end{supplement}


\printaddresses

\end{document}